\documentclass[%
reprint,
superscriptaddress,
 amsmath,amssymb,
 aps,
 prx,
]{revtex4-1}

\usepackage{natbib}
\usepackage{physics}
\usepackage{amsmath}
\usepackage{subfigure}
\usepackage[most]{tcolorbox}

\tcbuselibrary{theorems}

\newtcbtheorem[number within=section]{mybox}{Box}{
    halign title=left, 
    halign=left, 
}{bx}

\usepackage{enumitem}


\usepackage{graphicx}
\graphicspath{{figures/}}
\usepackage{booktabs}
\usepackage[colorlinks,linkcolor=blue,citecolor=blue,urlcolor=blue]{hyperref}

\newtheorem{theorem}{Theorem}

\begin{document}


\title{Quantum key distribution overcoming practical correlated intensity fluctuations}


\author{Jia-Xuan Li}\email{These authors contributed equally to this work}
\affiliation{CAS Key Laboratory of Quantum Information, University of Science and Technology of China, Hefei, Anhui 230026, China}
\affiliation{CAS Center for Excellence in Quantum Information and Quantum Physics, University of Science and Technology of China, Hefei, Anhui 230026, China}
\author{Feng-Yu Lu}\email{These authors contributed equally to this work}
\author{Ze-Hao Wang}\email{These authors contributed equally to this work}
\affiliation{CAS Key Laboratory of Quantum Information, University of Science and Technology of China, Hefei, Anhui 230026, China}
\affiliation{CAS Center for Excellence in Quantum Information and Quantum Physics, University of Science and Technology of China, Hefei, Anhui 230026, China}
\author{Víctor Zapatero}
\author{Marcos Curty}
\affiliation{Vigo Quantum Communication Center, University of Vigo, Vigo E-36310, Spain}
\affiliation{Escuela de Ingeniería de Telecomunicación, Department of Signal Theory and Communications, University of Vigo, Vigo E-36310, Spain}
\affiliation{AtlanTTic Research Center, University of Vigo, Vigo E-36310, Spain}
\author{Shuang Wang}\email{wshuang@ustc.edu.cn}
\author{Zhen-Qiang Yin}\email{yinzq@ustc.edu.cn}
\affiliation{CAS Key Laboratory of Quantum Information, University of Science and Technology of China, Hefei, Anhui 230026, China}
\affiliation{CAS Center for Excellence in Quantum Information and Quantum Physics, University of Science and Technology of China, Hefei, Anhui 230026, China}
\affiliation{Hefei National Laboratory, University of Science and Technology of China, Hefei 230088, China}
\author{Wei Chen}
\author{De-Yong He}
\author{Guang-Can Guo}
\author{Zheng-Fu Han}
\affiliation{CAS Key Laboratory of Quantum Information, University of Science and Technology of China, Hefei, Anhui 230026, China}
\affiliation{CAS Center for Excellence in Quantum Information and Quantum Physics, University of Science and Technology of China, Hefei, Anhui 230026, China}
\affiliation{Hefei National Laboratory, University of Science and Technology of China, Hefei 230088, China}


\date{\today}

\begin{abstract}
\textcolor{black}{
    Intensity correlations between neighboring pulses open a prevalent yet often overlooked security loophole in decoy-state quantum key distribution (QKD). 
    As a solution, we present 
    {and experimentally demonstrate} 
    an intensity-correlation-tolerant QKD protocol that mitigates the negative effect 
    that this phenomenon has on the secret key rate according to existing security analyses. 
    Compared to previous approaches, our method significantly enhances the robustness against correlations, 
    notably improving both the maximum transmission distances and the achievable secret key rates across different scenarios. 
        %
        %
        By relaxing constraints on correlation parameters, our protocol enables practical devices to counter intensity correlations.
        We experimentally demonstrate this first practical solution that directly overcomes this security vulnerability,
    establish the feasibility and efficacy of our proposal, 
    taking a major step towards loophole-free and high-performance QKD.
    }
\end{abstract}


\maketitle

\section{introduction}

Quantum key distribution \cite{BB84} (QKD) stands at the forefront of secure communication protocols, as it enables two remote users, Alice and Bob, to share secret keys whose security is guaranteed by the principles of quantum mechanics~\cite{doi:10.1126/science.283.5410.2050, PhysRevLett.85.441, RevModPhys.81.1301, doi:10.1142/S0219749908003256}. However, QKD security proofs typically extract secret bits from the raw data originating from single-photons, and on-demand high-quality single-photon sources at telecom wavelengths are not available yet. A popular solution to solve this pressing issue is the decoy-state method~\cite{PhysRevLett.91.057901, PhysRevLett.94.230503, PhysRevLett.94.230504}, which provides the same secret key rate scaling as single-photon sources by means of using laser sources emitting phase-randomized weak coherent pulses (PRWCPs). Indeed, this technique is a standard tool in current QKD implementations~\cite{Takesue2007,Lucamarini:13,Yuan:18,grunenfelder2023fast}. 

An important breakthrough direction for advancing 
QKD is to increase its secret key rate, for which fast operating QKD systems are being developed
\cite{Takesue2007, Lucamarini:13, Yuan:18, grunenfelder2023fast,doi:10.1126sciadv.1701491, PhysRevLett.121.190502, Grnenfelder2020, Li2023}. However, due to memory effects in the devices and the electronics that control them, high-clock-rate decoy-state QKD systems face a troublesome implementation security problem: 
intensity correlations~\cite{Grnenfelder2020, PhysRevA.90.032320, Roberts:18, Yoshino2018, Lu2021, 10050030,zapatero2021security, PhysRevApplied.18.044069}. 
This means that the intensity setting of any given round may influence the actual intensity emitted in subsequent rounds, 
resulting in a partial distinguishability of the intensity settings. 
This breaks a core assumption of the decoy-state method, 
posing an underestimated threat to the security of QKD~\cite{zapatero2021security, PhysRevApplied.18.044069,doi:10.1126/sciadv.aaz4487}. 
To address this problem, various security analyses have been proposed \cite{Yoshino2018, zapatero2021security, PhysRevApplied.18.044069}. 
However, practical devices \cite{9907821, PhysRevApplied.19.014048, Xie:19} often struggle to meet the criteria set by these analyses (such as the magnitude of the correlations) leading to low or even vanishing secret key rates.

Here, we solve this crucial limitation by proposing an approach that we name intensity-correlation-tolerant QKD, which is capable of mitigating the intensity correlations problem in QKD devices. By adding a local monitor, our protocol enables common devices to achieve notably higher secret key rates and longer transmission distances than previous solutions in the presence of this type of correlations. Importantly, we experimentally demonstrate the feasibility and effectiveness of our approach. This advancement is a significant step towards loophole-free and high-performance QKD.

\begin{figure*}
    \begin{mybox}{Protocol description}{ict-1}
        \begin{enumerate}[label=\arabic*., leftmargin=*]
            \item 
            \textbf{State preparation:}
            In the $k$-th round ($k=1,2,\ldots,N$), Alice selects an intensity setting $a_k \in A = \{\mu, \nu, \omega\}$ with probability $p_{a_k}$, \textcolor{black}{and a bit-and-basis setting $r_k \in R = \{Z0, Z1 , X0, X1\}$ with probability $p_{r_k}$, where $p_{Z0} = p_{Z1} = q_Z / 2$ and $p_{X0} = p_{X1} = q_X / 2$ for $q_Z+q_X=1$.}
            \textcolor{black}{Then, she prepares a PRWCP accordingly}, and a \textbf{local monitor} measures a fraction of the generated PRWCP with a photodetector with relative efficiency $\eta_{\rm m}$. 
            \item 
            \textbf{Measurement:}
            Bob \textcolor{black}{measures the received states, and announces the successful measurement results}. Alice and Bob record these later rounds and their corresponding bases and raw key bits. \textcolor{black}{Also, Alice stores the intensity settings for all rounds.}
            \item 
            \textbf{Sifting:}
            \textcolor{black}{Alice and Bob broadcast their basis choices for each round. They keep the raw key bits for the $Z(X)$-basis coincidences to form the sifted key (parameter estimation data).}
            \item 
            \textbf{Parameter estimation:}
            \textcolor{black}{Bob discloses his measurement outcomes for the parameter estimation data.} Alice classifies the results recorded by her local monitor in the state preparation step \textcolor{black}{into $3^{\xi+1}$ categories ---matching the possible records $\{k, \xi\}$--- and computes the \textbf{correlation parameters} $\tau^{\xi}_{ a_{k}, a_{k}' , n}$. With this information, she uses the \textbf{enhanced decoy-state method} to estimate the $Z$ and $X$ basis single-photon yields, together with the phase error rate in the $Z$ basis.}
            \item 
            \textbf{Key distillation:}
            Alice and Bob perform error correction and privacy amplification based on the results of the parameter estimation step, and then use the data in the $Z$ basis to generate the secret keys.
        \end{enumerate}
    \end{mybox}
\end{figure*}
\section{Assumptions\label{sec:Assumptions}}

    To characterize intensity correlations and fluctuations, we assume a general model from~\cite{zapatero2021security}. 
    Precisely, in any given round $k$, the correlations do not compromise the Poissonian character of the photon-number statistics 
    of
    the source conditioned on the value of the actual intensity, $\alpha_k$. Nevertheless, the latter does not match the selected intensity setting, say $a_{k}$, 
    but it is also influenced by the previous $\xi$ settings, $a_{k-\xi}a_{k-\xi+1}\ldots a_{k-1}$, 
    for a certain correlation range $\xi$. That is, we model $\alpha_k$ as a random variable, 
    in such a way that every possible record $\{k,\xi\}:=a_{k-\xi}a_{k-\xi+1}\ldots a_{k}$ determines a conditional probability distribution 
    for $\alpha_k$, with expectation $\bar{\alpha}_{\{k,\xi\}}$. 
    Here, we describe this randomness using the relative deviation $\delta_{\{k,\xi\}}$ ---such that $\alpha_k = \bar{\alpha}_{\{k,\xi\}} (1+\delta_{\{k,\xi\}})$--- 
    and assume that the probability density function of $\delta_{\{k,\xi\}}$, $g_{\{k,\xi\}}$, 
    is only nonzero in a certain record-dependent interval $[\delta_{\{k,\xi\}}^{-},\delta_{\{k,\xi\}}^{+}]$. 
    For later convenience, we define $\delta_{{\rm max},a_k}^{\rm rand}=\max_{\{k,\xi\}}\{|\delta_{\{k,\xi\}}^{\pm}|\}$ 
    and $\delta_{{\rm max},a_k}^{\rm corr}=\max_{\{k,\xi\}}\{|1-\bar{\alpha}_{\{k,\xi\}}/a_{k}|\}$,
    which denote the the magnitude of sequence-dependent random fluctuations 
    and the size of correlation of the average intensity, respectively.

We make the following assumptions on the intensity correlations and fluctuations.

{\bf Assumption 1:}
The intensity correlations do not compromise the Poissonian character of the photon-number statistics of the source 
\cite{grunenfelder2020performance,yoshino2018quantum,lu2021intensity,kang2023patterning}.
That is to say, given the actual intensity $\alpha_k$ prepared in the $k$-th round, the conditional photon-number statistics satisfy
\begin{equation}
    \label{equ:pn_alpha}
        \begin{aligned}
            p_{n}|_{\alpha_k}
            =
            \frac{
                {\rm e}^{-\alpha_k}(\alpha_k)^n
            }
            {n!}.
        \end{aligned} 
\end{equation}

{\bf Assumption 2:}
The intensity correlations have a finite range $\xi$, meaning that the intensity setting $a_{k-i}$ of the $(k-i)$-th round does not influence $\alpha_k$ if $i > \xi$.

{\bf Assumption 3:}
Let us introduce the shorthand notation $\{i,j\}$ to describe the record of settings $a_{i-j} a_{i-j+1}...a_{i}$. Each record $\{k,\xi\}$ determines a conditional probability distribution for $\alpha_k$, with expectation $\bar{\alpha}_{\{k,\xi\}}$. For convenience, we describe this randomness using the relative deviation $\delta_{\{k,\xi\}}$, such that $\alpha_k = \bar{\alpha}_{\{k,\xi\}} (1+\delta_{\{k,\xi\}})$. The probability density function of $\delta_{\{k,\xi\}}$, which we denote by $g_{\{k,\xi\}}$, is only nonzero in the record-dependent interval $[\delta_{\{k,\xi\}}^{-},\delta_{\{k,\xi\}}^{+}]$.\\

We remark that, by definition,
    \begin{equation}
        \label{equ:note514}
            \begin{aligned}
                \int_{\delta_{\{k,\xi\}}^{-}}^{\delta_{\{k,\xi\}}^{+}}  
                {g}_{\{k,\xi\}}(\delta)
                \delta
                {\rm d}\delta
                =0.
            \end{aligned}
    \end{equation}

Also, for a given record $\{k,\xi\}$, the conditional photon-number statistics of round $k$ satisfy
\begin{widetext}
\begin{equation}
    \label{equ:pn_all}
        \begin{aligned}
            p_{n}|_{\{k,\xi\}}
            =
            \int_{\delta_{\{k,\xi\}}^{-}}^{\delta_{\{k,\xi\}}^{+}}  
            {g}_{\{k,\xi\}}(\delta)
                {\rm exp}\left({- \bar{\alpha}_{\{k,\xi\}}(1+\delta)}\right)
            \frac{
                [
                    { \bar{\alpha}_{\{k,\xi\}}(1+\delta)}
                ]^n
            }{
                n!
            }
            {\rm d}\delta
        \end{aligned}
    .
\end{equation}
\end{widetext}

Finally, matching the notation in the main text, we introduce $\delta_{{\rm max},a_k}^{\rm rand}=\max_{\{k,\xi\}}\{|\delta_{\{k,\xi\}}^{\pm}|\}$ and $\delta_{{\rm max},a_k}^{\rm corr}=\max_{\{k,\xi\}}\{|1-\bar{\alpha}_{\{k,\xi\}}/a_{k}|\}$.

\section{enhanced decoy-state method}
\label{sec:enhanced decoy-state method}

Importantly, in the presence of intensity correlations, the $n$-photon yield and error rate associated to different intensity settings can be distinct. 
    To address this issue, we introduce an enhanced decoy-state method. 
    Crucially, we consider finer-grained decoy-state constraints by grouping the measurement counts 
    at Bob's receiver according to the record of settings rather than using the last setting alone as in standard analyses. 
    Specifically, our protocol involves imposing constraints in two key aspects. 
    The first is a photon number constraint. 
    Similar to the standard decoy-state method, we truncate the photon number at $n_{\rm cut}$ to establish upper and lower bounds. 
    However, due to security concerns arising from correlations, we need to classify the detection rates and error rates according to the sequence 
    $\{k,\xi\}$. Despite this, the constraints alone are insufficient for parameter estimation, 
    which leads us to introduce a second constraint. 
    Specifically, we employ a mathematical tool called the Cauchy-Schwarz (CS) constraint 
    \cite{zapatero2021security, PhysRevApplied.18.044069} to set limits on the bias in detection statistics associated with different intensities.



\subsection{The CS constraint}
\label{Sec:CSbudengs1}

As originally observed in~\cite{zapatero2021security}, in the presence of intensity correlations, the yields and error probabilities in any given round may differ for different records of settings. As a consequence, the decoy-state method~\cite{PhysRevLett.91.057901,PhysRevLett.94.230503,PhysRevLett.94.230504} alone does not enable a tight parameter estimation, and additional constraints are required. In accordance with \cite{zapatero2021security}, we address this problem by using the so-called Cauchy-Schwarz (CS) constraints, a tool previously exploited in~\cite{doi:10.1126/sciadv.aaz4487,PhysRevApplied.18.044069} too. Specifically, the CS constraint restricts the bias between the measurement statistics of two different quantum states 
when subject to the same measurement.
The CS constraint can be stated as follows.
\begin{theorem}\label{theorem 1}
Let $\ket{a}$ and $\ket{b}$ be two pure states of an arbitrary Hilbert space $\mathcal{H}$. For any operator $\widehat{O}$ on $\mathcal{H}$ such that $0\leq \widehat{O} \leq 1$,
    \begin{equation}
        \label{equ:CS-BDS-1}
            \begin{aligned}
                G_-\left(
                    {\rm Tr}\left[
                        \widehat{O} \ket{a} \bra{a}
                    \right],
                    \left\lvert
                        \braket{a}{b}
                    \right\rvert^2
                \right)
                \leq
                {\rm Tr}\left[
                    \widehat{O} \ket{b} \bra{b}
                \right]
                \\
                \leq
                G_+\left(
                    {\rm Tr}\left[
                        \widehat{O} \ket{a} \bra{a}
                    \right],
                    \left\lvert
                        \braket{a}{b}
                    \right\rvert ^2
                \right)
            \end{aligned}
        ,
    \end{equation}
    where
    \begin{equation}
        \label{equ:Gpm}
            \begin{aligned}
                G_- &=
                \begin{cases}
                    g_-(x,y) \quad x>1-y
                    \\
                    0 \quad \, \ \ \qquad x\leq 1-y
                \end{cases}
                ,
                \\
                G_+ &=
                \begin{cases}
                    1 \quad \, \ \ \qquad x\geq y
                    \\
                    g_+(x,y) \quad x<y
                \end{cases}
                ,
            \end{aligned}
    \end{equation}
    with
    $g_{\pm}(x,y) = x + (1-2x)(1-y) \pm 2\sqrt{x(1-x)y(1-y)}$.

    Proof: See the Supplementary Materials of Ref.\cite{doi:10.1126/sciadv.aaz4487}.
\end{theorem}

Essentially, Eq.~(\ref{equ:CS-BDS-1}) allows to set quantitative bounds on the detection statistics arising from different records of settings. However, because of their non-linearity, these bounds cannot be directly plugged into decoy-state linear programs. Notwithstanding, in virtue of the convexity of the $G_{\pm}(x,y)$ functions, suitable linearizations of the CS constraints follow. In particular, for any reference value $c \in [0,1]$, we have
\begin{equation}
    \label{equ:CS-BDS-XXBB}
        \begin{aligned}
            &
            G_-\left(
                c,
                \left\lvert
                    \braket{a}{b}
                \right\rvert ^2
            \right)
            +
            G_-'\left(
                c,
                \left\lvert
                    \braket{a}{b}
                \right\rvert ^2
            \right)
            \left(
                {\rm Tr}\left[
                    \widehat{O} \ket{a} \bra{a}
                \right]
                -c
            \right)
            \\
            \leq
            &
            {\rm Tr}\left[
                \widehat{O} \ket{b} \bra{b}
            \right]
            \\
            \leq
            &
            G_+\left(
                c,
                \left\lvert
                    \braket{a}{b}
                \right\rvert ^2
            \right)
            +
            G_+'\left(
                c,
                \left\lvert
                    \braket{a}{b}
                \right\rvert ^2
            \right)
            \left(
                {\rm Tr}\left[
                    \widehat{O} \ket{a} \bra{a}
                \right]
                -c
            \right)
        \end{aligned}
    ,
\end{equation}
where
\begin{equation}
    \label{equ:Gpm-xxbb}
        \begin{aligned}
            G_-' &=
            \begin{cases}
                g_-'(x,y) \quad x>1-y
                \\
                0 \quad \, \ \ \qquad x\leq 1-y
            \end{cases}
            ,
            \\
            G_+' &=
            \begin{cases}
                0 \quad \, \ \ \qquad x\geq y
                \\
                g_+'(x,y) \quad x<y
            \end{cases}
            ,
        \end{aligned}
\end{equation}
with
$g_{\pm}'(x,y) = -1 +2y \pm (1-2x) \sqrt{y(1-y)/x(1-x)}$. Obviously, although the value of $c$ can be set arbitrarily, it has a direct impact on the tightness of the constraints. In this regard, our preferred choices for these reference parameters are discussed in Section~\ref{sec:sdsfvsdaa}.

\subsection{Intensity correlation parameter}

One of the key ideas for addressing intensity correlations is that,
in the context of general attacks, 
setting CS constraints on the measurement statistics of a fixed round $k$ requires to compute the inner product between specific quantum states across all $N$ protocol rounds~\cite{zapatero2021security,PhysRevApplied.18.044069}
(although respectively conditioned on the two records of settings whose statistics are to be compared). 
In this subsection, we outline the calculation of these inner products, and for this purpose we restore the explicit notation of the intensity-setting subscripts for clarity. For further technical explanations, the reader is referred to~\cite{zapatero2021security}.

In the entanglement-based picture, the global input state of all protocol rounds can be described as
\begin{widetext}
\begin{equation}
    \label{equ:NroundPSIall}
    \begin{aligned}
        \ket{\Psi}= \left[ \sum_{a_{1}\ldots{}a_{N}} \sum_{r_{1}\ldots{}r_{N}}  
        \left( 
            \prod_{i=1}^N \sqrt{ p_{a_i}q_{r_i}} 
        \right) 
                    \left(
                        \bigotimes_{i=1}^N \ket{a_i}_{A_i}   \ket{r_i}_{{A_i}'}
                        \ket{\psi_{a_{1}\ldots{}a_{i}}^{r_i}}_{B_i C_i}
                    \right)
                    \right]  
                    \otimes \ket{0}_E
        ,
    \end{aligned}
\end{equation}
\end{widetext}
where 
$\ket{a_i}_{A_i}$ is a virtual ancilla storing the intensity setting in round $i$,
$ \ket{r_i}_{{A_i}'}$ is a virtual ancilla storing the encoded BB84 state in round $i$ (\textit{i.e.}, the bit and basis information),
$p_{a_i}$ is the probability of choosing the intensity setting $a_{i}$ in round $i$, $q_{r_i}$ is the probability of choosing the BB84 state $r_{i}$ in round $i$, and 
$\ket{0}_{E}$ is the vacuum state of Eve's system. Also, for any given round $i$, we have defined
\begin{equation}
    \label{equ:iRoudPsi}
    \begin{aligned}
        \ket{\psi_{a_{1}\ldots{}a_{i}}^{r_i}}_{B_i C_i}
        =
        \sum_{n_i = 0}^{\infty}  \sqrt{p_{n_i}|_{a_{1}\ldots{}a_{i}}}
        \ket{t_{n_i}}_{C_i}
        \ket{n_i^{r_i}}_{B_i}
        ,
    \end{aligned}
\end{equation}
where $p_{n_i}|_{a_{1}\ldots{}a_{i}}$ is the conditional $n_{i}$-photon probability given the record $a_{1}\ldots{}a_{i}$, $\ket{t_{n_i}}_{C_i}$ is a virtual ancilla storing the photon number ${n_i}$, and $\ket{n_i^{r_i}}_{B_i}$ is a Fock state with $n_{i}$ photons encoding the BB84 state $r_{i}$. Note that, given the finite range $\xi$ of the correlations, for all $i>\xi$, we can replace $p_{n_i}|_{a_{1}\ldots{}a_{i}}$ by $p_{n_i}|_{a_{i-\xi}\ldots{}a_{i}}$ and $\ket{\psi_{a_{1}\ldots{}a_{i}}^{r_i}}_{B_i C_i}$ by $\ket{\psi_{a_{i-\xi}\ldots{}a_{i}}^{r_i}}_{B_i C_i}$.

In principle, for any given round $k$, one could establish CS constraints between the detection statistics of any two arbitrary records of settings. Nevertheless, this exhaustive approach would result in a number of CS constraints that increases exponentially with the correlation range, which is computationally prohibitive. Instead, we follow the simpler analysis presented in~\cite{PhysRevApplied.18.044069}, where only the tightest CS constraints are considered. Particularly, for round $k$, one only relates the yields/error yields of records of intensity settings exclusively differing in $a_{k}$. Let $a_{k-\xi}\ldots{}a_{k-1}a_k$ and $a_{k-\xi}\ldots{}a_{k-1}a_k'$ be any two such records. As shown in~\cite{zapatero2021security,PhysRevApplied.18.044069}, it turns out that the relevant overlap to compute in order to evaluate the CS constraints between these two records is
\begin{equation}
\left\lvert 
    \bra{{\psi}_{a_{k-\xi}\ldots{}a_{k-1}a_k,n}}
    \ket{{\psi}_{a_{k-\xi}\ldots{}a_{k-1}a_k',n}}
\right\rvert,
\end{equation}
where
\begin{equation}
\ket{{\psi}_{a_{k-\xi}\ldots{}a_{k-1}a_k,n}}=\displaystyle{\frac{\ket{\widetilde{\psi}_{a_{k-\xi}\ldots{}a_{k-1}a_k,n}}}{\left\lVert{\ket{\widetilde{\psi}_{a_{k-\xi}\ldots{}a_{k-1}a_k,n}}}\right\rVert}},
\end{equation}
for $\ket{\widetilde{\psi}_{a_{k-\xi}\ldots{}a_{k-1}a_k,n}}=\bra{a_{k-\xi}}_{A_{k-\xi}}\otimes\ldots\otimes\bra{a_{k-1}}_{A_{k-1}}\otimes\bra{a_{k}}_{A_{k}}\otimes\bra{t_n}_{C_k}\ket{\Psi}$. Explicit calculation of this overlap follows identically as in~\cite{PhysRevApplied.18.044069} and yields
\begin{widetext}
\begin{equation}\label{equ:inner1}
\left\lvert \bra{{\psi}_{a_{k-\xi}\ldots{}a_{k-1}a_k,n}}\ket{{\psi_{a_{k-\xi}\ldots{}a_{k-1}a_k',n}}}\right\rvert=\sum_{a_{k+1}\in A}\ldots\sum_{a_{k+\xi}\in A}\left(\prod_{i=k+1}^{k+\xi} p_{a_i}\sum_{m=0}^{\infty}\sqrt{p_{m}|_{a_{i-\xi}\ldots{}a_{k}\ldots{}a_{i}}p_{m}|_{a_{i-\xi}\ldots{}a_{k}'\ldots{}a_{i}}}\right).
\end{equation}
As one would expect, the overlap is dependent on the correlation function $g_{\{k-1,\xi-1\}}$ through the conditional photon-number statistics. Since $g_{\{k-1,\xi-1\}}$ is unknown, one must derive correlation-function-independent lower bounds on the photon-number statistics ---say, $p_{n}^{\rm L}|_{a_{k-\xi}\ldots{}a_{k}}$ for the specific record $a_{k-\xi}\ldots{}a_{k}$--- in order to possibly lower bound the overlap in Eq.~(\ref{equ:inner1}). In this respect, we recall that lower bounds on the overlaps are necessary to reach loosened but correlation-function-independent CS constraints~\cite{PhysRevApplied.18.044069}. In fact, such loosened CS constraints rely on lower bounds on the squared overlaps, which we refer to as the intensity correlation parameters,
\begin{equation}
    \label{equ:tauChongXie}
    \begin{aligned}
        \tau^{\xi}_{ a_{k}, a_{k}' , n}
        :=
        \left\{\sum_{a_{k+1}\in A}\ldots\sum_{a_{k+\xi}\in A}\left(\prod_{i=k+1}^{k+\xi} p_{a_i}\sum_{m=0}^{\infty}\sqrt{p_{m}^{\rm L}|_{a_{i-\xi}\ldots{}a_{k}\ldots{}a_{i}}p_{m}^{\rm L}|_{a_{i-\xi}\ldots{}a_{k}'\ldots{}a_{i}}}\right)\right\}^2
    \end{aligned},
\end{equation}
\end{widetext}
for any given round $k$, any pair of distinct settings $a_{k}$ and $a_{k'}$, and any photon number $n$. The intensity correlation parameters are calculated in Sec.~\ref{sec asdsfhakscfaeracb}.

\subsection{Enhanced linear programs}
In this subsection, we write down the linear programs required to estimate the asymptotic secret key rate. In the first place, the standard decoy-state constraints can be stated in the form
\begin{equation}
    \label{equ:tiaojiankaishi}
    \begin{aligned}
        \frac{ 
                    Z_{\{k,\xi\}} 
            }{
                q_Z^2 \prod_{i=k-\xi}^{k} p_{a_i}
            }
        =
        \sum_{n=0}^{\infty}
            p_{n}|_{\{k,\xi\}}
            y_{n,\{k,\xi\}}
    \end{aligned}
    ,
\end{equation}
where the left-hand side is the probability of a click occurring in round $k$ conditioned on a $Z$-basis match and the record of settings being $\{k,\xi\}$, and $y_{n,\{k,\xi\}}$ denotes the corresponding $n$-photon yield. As is customary in decoy-state analyses, one can select a threshold photon-number $n_{\rm cut}$ in order to split Eq.~(\ref{equ:tiaojiankaishi}) into two complementary bounds,
\begin{widetext}
\begin{equation}
    \label{gains}
    \begin{aligned}
        &
            \frac{
                    Z_{\{k,\xi\}} 
            }{
                q_Z^2 \prod_{i=k-\xi}^{k} p_{a_i}
            }
            \geq
            \sum_{n=0}^{n_{\rm cut}}
            p_{n}^{\rm L}|_{\{k,\xi\}}
            y_{n,\{k,\xi\}}
            \quad
            (
                a_{k-\xi},\ldots,a_{k} \in A
            )
            ,
            \\
            &
            \frac{
                    Z_{\{k,\xi\}}
            }{
                q_Z^2 \prod_{i=k-\xi}^{k} p_{a_i}
            }
            \leq
            \sum_{n=0}^{n_{\rm cut}}
            p_{n}^{\rm U}|_{\{k,\xi\}}
            y_{n,\{k,\xi\}}
            +
            1-\sum_{n=0}^{n_{\rm cut}}
            p_{n}^{\rm L}|_{\{k,\xi\}}
            \quad
            (
                a_{k-\xi},\ldots,a_{k} \in A
            )
            .
    \end{aligned}
\end{equation}
Analogously, for the error statistics (say, of the $X$ basis) we have
\begin{equation}
    \label{error_gains}
    \begin{aligned}
        &
            \frac{
                    E_{\{k,\xi\}}
            }{
                q_X^2 \prod_{i=k-\xi}^{k} p_{a_i}
            }
            \geq
            \sum_{n=0}^{n_{\rm cut}}
            p_{n}^{\rm L}|_{\{k,\xi\}}
            h_{n,\{k,\xi\}}
            \quad
            (
                a_{k-\xi},\ldots,a_{k} \in A
            )
            ,
            \\
            &
            \frac{
                    E_{\{k,\xi\}}
            }{
                q_X^2 \prod_{i=k-\xi}^{k} p_{a_i}
            }
            \leq
            \sum_{n=0}^{n_{\rm cut}}
            p_{n}^{\rm U}|_{\{k,\xi\}}
            h_{n,\{k,\xi\}}
            +
            1-\sum_{n=0}^{n_{\rm cut}}
            p_{n}^{\rm L}|_{\{k,\xi\}}
            \quad
            (
                a_{k-\xi},\ldots,a_{k} \in A
            )
            ,
    \end{aligned}
\end{equation}
where $E_{\{k,\xi\}}\bigr/\bigl(q_X^2 \prod_{i=k-\xi}^{k} p_{a_i}\bigr)$ is the probability of a click and a bit error occurring in round $k$ conditioned on a $X$-basis match and the record of settings being $\{k,\xi\}$, and $h_{n,\{k,\xi\}}$ denotes the corresponding $n$-photon error yield.

On the other hand, the linearized CS constraints that arise from Eq.~(\ref{equ:CS-BDS-XXBB}), when applied to the yields, can be written as~\cite{zapatero2021security,PhysRevApplied.18.044069}
\begin{equation}
        \label{CS_yields}
        \begin{aligned}
            &
            c^{+}_{\{k-1,\xi-1\},  a_{k},a_{k}',n} + m^{+}_{\{k-1,\xi-1\},  a_{k},a_{k}',n} y_{n,\{k-1,\xi-1\} a_{k}}
            \geq
            y_{n,\{k-1,\xi-1\} a_{k}'}
            \quad
            (
                a_{k-\xi},\ldots,a_{k},a_{k}' \in A
                ;
                a_{k} \neq a_{k}
                ;
                n = 0 , \ldots , n_{\rm cut}
            )
            ,
            \\
            &
            c^{-}_{\{k-1,\xi-1\},  a_{k},a_{k}',n} + m^{-}_{\{k-1,\xi-1\},  a_{k},a_{k}',n} y_{n,\{k-1,\xi-1\} a_{k}}
            \geq
            y_{n,\{k-1,\xi-1\} a_{k}'}
            \quad
            (
                a_{k-\xi},\ldots,a_{k},a_{k}' \in A
                ;
                a_{k} \neq a_{k}
                ;
                n = 0 , \ldots , n_{\rm cut}
            )
            ,
        \end{aligned}
    \end{equation}
with
\begin{equation}
        \label{equ:cmts}
        \begin{aligned}
            c^{\pm}_{\{k-1,\xi-1\},  a_{k},a_{k}',n}
            =
            &
            G_{\pm}(
                \widetilde{y}_{n,\{k-1,\xi-1\} a_{k}}
                ,
                \tau^{\xi}_{ a_{k}, a_{k}' , n}
            )
            -
            G_{\pm}'(
                \widetilde{y}_{n,\{k-1,\xi-1\} a_{k}}
                ,
                \tau^{\xi}_{ a_{k}, a_{k}' , n}
            )
            \widetilde{y}_{n,\{k-1,\xi-1\} a_{k}}
            ,
            \\
            m^{\pm}_{\{k-1,\xi-1\},  a_{k},a_{k}',n}
            =
            &
            G_{\pm}'(
                \widetilde{y}_{n,\{k-1,\xi-1\} a_{k}}
                ,
                \tau^{\xi}_{ a_{k}, a_{k}' , n}
            )
            ,
        \end{aligned}
    \end{equation}
    where the reference values $\{\widetilde{y}_{n,\{k,\xi\}}\}$ of the linearization are provided in Section~\ref{sec:sdsfvsdaa}.

    In a similar fashion, the resulting linearized CS constraints for the error statistics read
    \begin{equation}
        \label{CS_error_yields}
        \begin{aligned}
           &
            t^{+}_{\{k-1,\xi-1\},  a_{k},a_{k}',n} + s^{+}_{\{k-1,\xi-1\},  a_{k},a_{k}',n} h_{n,\{k-1,\xi-1\} a_{k}}
            \geq
            h_{n,\{k-1,\xi-1\} a_{k}'}
            \quad
            (
                a_{k-\xi},\ldots,a_{k},a_{k}' \in A
                ;
                a_{k} \neq a_{k}
                ;
                n = 0 , \ldots , n_{\rm cut}
            )
            ,
            \\
            &
            t^{-}_{\{k-1,\xi-1\},  a_{k},a_{k}',n} + s^{-}_{\{k-1,\xi-1\},  a_{k},a_{k}',n} h_{n,\{k-1,\xi-1\} a_{k}}
            \geq
            h_{n,\{k-1,\xi-1\} a_{k}'}
            \quad
            (
                a_{k-\xi},\ldots,a_{k},a_{k}' \in A
                ;
                a_{k} \neq a_{k}
                ;
                n = 0 , \ldots , n_{\rm cut}
            )
            ,
        \end{aligned}
    \end{equation}
    for
    \begin{equation}
        \label{equ:cmts}
        \begin{aligned}
            t^{\pm}_{\{k-1,\xi-1\},  a_{k},a_{k}',n}
            =
            &
            G_{\pm}(
                \widetilde{h}_{n,\{k-1,\xi-1\} a_{k}}
                ,
                \tau^{\xi}_{ a_{k}, a_{k}' , n}
            )
            -
            G_{\pm}'(
                \widetilde{h}_{n,\{k-1,\xi-1\} a_{k}}
                ,
                \tau^{\xi}_{ a_{k}, a_{k}' , n}
            )
            \widetilde{h}_{n,\{k-1,\xi-1\} a_{k}}
            ,
            \\
            s^{\pm}_{\{k-1,\xi-1\},  a_{k},a_{k}',n}
            =
            &
            G_{\pm}'(
                \widetilde{h}_{n,\{k-1,\xi-1\} a_{k}}
                ,
                \tau^{\xi}_{ a_{k}, a_{k}' , n}
            )
            ,
        \end{aligned}
    \end{equation}
    where the reference values $\{\widetilde{h}_{n,\{k,\xi\}}\}$ are given in Section~\ref{sec:sdsfvsdaa}.
    
    Putting it all together, Eq.~(\ref{gains}) and Eq.~(\ref{CS_yields}) plus the boundary constraints $0\leq{}y_{n,\{k,\xi\}}\leq{}1$ compose the round-dependent linear program of the $Z$-basis detection statistics, whose objective function is given by
    \begin{equation}\label{target_yields}
    Z_{1,\mu}^{(k)\mathrm{L}}=q_{\rm Z}^{2}p_{\mu}\sum_{a_{k-\xi}\in{}A}\ldots{}\sum_{a_{k-1}\in{}A}\left(\prod_{i=k-\xi}^{k-1}p_{a_{i}}\right)p_{1}^{\rm L}|_{\{k-1,\xi-1\}\mu}y_{1,\{k-1,\xi-1\}\mu}.
    \end{equation}
    Namely, a lower bound on the total probability of a $Z$-basis match and a signal-setting single-photon click occurring in round $k$ (averaged over all possible records $\{k-1,\xi-1\}$).
    
    Importantly, an equivalent LP for the $X$-basis detection statistics is obtained by replacing $X$ by $Z$ where convenient. The objective function of such LP, defined analogously as $Z_{1,\mu}^{(k)\mathrm{L}}$ but for the $X$ basis, is denoted as $X_{1,\mu}^{(k)\mathrm{L}}$.
    
    Analogously, Eq.~(\ref{error_gains}) and Eq.~(\ref{CS_error_yields}) plus the boundary constraints $0\leq{}h_{n,\{k,\xi\}}\leq{}1$ compose the round-dependent linear program of the $X$-basis error statistics, whose objective function is given by
    \begin{equation}\label{target_error_yields}
    E_{1,\mu}^{(k)\mathrm{L}}=q_{\rm Z}^{2}p_{\mu}\sum_{a_{k-\xi}\in{}A}\ldots{}\sum_{a_{k-1}\in{}A}\left(\prod_{i=k-\xi}^{k-1}p_{a_{i}}\right)p_{1}^{\rm U}|_{\{k-1,\xi-1\}\mu}h_{1,\{k-1,\xi-1\}\mu}.
    \end{equation}
\end{widetext}
    Namely, an upper bound on the total probability of an $X$-basis match and a signal-setting single-photon bit error occurring in round $k$ (averaged over all possible records $\{k-1,\xi-1\}$).

To finish with, the relevant round-independent linear programs follow by summing over all $N$ protocol rounds 
and dividing over $N$ in both the objective functions and the constraints.

\subsection{Reference values for the linear CS constraints}
\label{sec:sdsfvsdaa}

\begin{figure*}[htbp]
    \centering
        \includegraphics[width=0.7\textwidth]{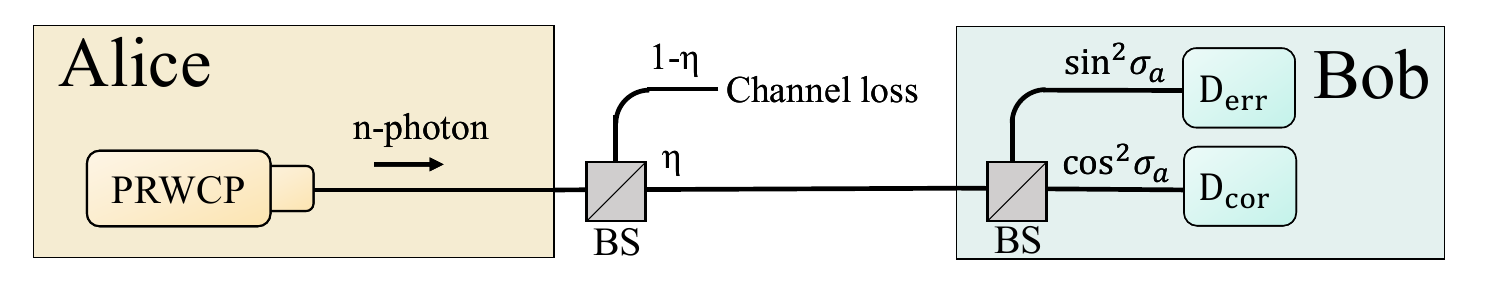}
    \caption{
                Illustration of the channel model that we use to compute the reference parameters for the CS constraints. PRWCP, phase-randomized weak coherent pulse;
                BS, beam splitter;
                $\eta$, overall transmittance (including channel transmittance and detection efficiency);
                $\sigma_a$, misalignment;
                $\rm D_{err}$, detector associated to the ``error" clicks;
                $\rm D_{cor}$, detector associated to the ``correct" (or ``no error") clicks.
            }
    \label{fig channel moud}
\end{figure*}

In this subsection we present the reference parameters $\widetilde{y}_{n,\{k-1,\xi-1\} a_{k}}$ and $\widetilde{h}_{n,\{k-1,\xi-1\} a_{k}}$ that we utilize in the CS constraints. 
As mentioned earlier, although these parameters can be chosen arbitrarily, they have an impact on the tightness of the CS constraints. 
In accordance with~\cite{zapatero2021security}, here we select them following a rather typical channel and detection model free of intensity correlations. 
The model is depicted in Fig.~\ref{fig channel moud}, and the main elements involved are described in the caption.

The reference $n$-photon yield in the model is simply given by
\begin{equation}
    \label{equ:yield-ref}
    \begin{aligned}
        \widetilde{y}_{n,\{k-1,\xi-1\} a_{k}}
        =
        1-(1-p_d)^2(1-\eta)^n
    \end{aligned}
    ,
\end{equation}
where $p_d$ is the dark count probability of Bob's detectors and $\eta$ is the overall transmittance of the system.

In order to write down the reference $n$-photon error probabilities, we average over all four combinations of genuine detection events (\textit{i.e.}, disregarding dark counts for the moment). These have associated probabilities $p_1=(1-\eta)^{n}$ (no detector clicks), $p_2 = (1 - \cos^2 \sigma_a\eta)^n - p_1$ (only $\rm D_{cor}$ clicks), $p_3 = (1 - \sin^2 \sigma_a\eta)^n - p_1$ (only $\rm D_{err}$ clicks) and $p_4 = 1 - p_1 - p_2 -p_3$ (both detectors click). Assuming that double clicks are randomly assigned to a specific detection outcome, one can readily identify the specific dark count combinations that map each of these four events to a bit error. In doing so, one obtains the following conditional error probabilities for all four events: $p_1^e = p_d(1-p_d) + \frac12 p_d^2$,
$p_2^e = (1-p_d)^2 + \frac32p_d(1-p_d) + \frac12 p_d^2$,
$p_3^e = \frac12p_d(1-p_d) + \frac12 p_d^2$ and
$p_4^e = \frac12(1-p_d)^2 + p_d(1-p_d) + \frac12 p_d^2$, and therefore we conclude that
\begin{equation}
    \label{equ:yield-ref}
    \begin{aligned}
        \widetilde{h}_{n,\{k-1,\xi-1\} a_{k}}
        =
        \sum_{i=1}^4
        p_i p_i^e
    \end{aligned}
    .
\end{equation}

\section{calculation of correlation parameters}
\label{sec:asdcqwecsg333}

    On the practical side, there are two challenges in implementing the enhanced decoy-state method. 
    First, the photon number constraint requires to accurately estimate tight upper and lower bounds on the photon number statistics. 
    Second, the CS constraints 
    require to determine lower bounds on the squared inner product between suitable quantum states that only differ on the last intensity setting,
    %
    which we have defined as $\tau^{\xi}_{ a_{k}, a_{k}' , n}$ in Eq.~(\ref{equ:tauChongXie}).
As our solution,
by including a local monitor as an integral component of the protocol, Alice can obtain a tight estimation of these parameters. 
Precisely, she measures a fraction of every PRWCP she sends to the quantum channel.
By categorizing the measurement results according to the setting sequences $\{ k, \xi \}$, Alice obtains the gain for each category. 
    Using this information, together with the accurate relative efficiency $\eta_{\rm m}$ between the pulse entering the channel
    and the pulse entering the detection module in the local monitor, she can 
    estimate the different photon-number probabilities for each category, further determining $\tau^{\xi}_{ a_{k}, a_{k}' , n}$.

In this section, we elucidate how our protocol calculates these parameters by means of using a local monitor situated at Alice's side
\cite{PhysRevA.75.052301,10050030}. Remarkably, if compared with the existing analyses~\cite{zapatero2021security,PhysRevApplied.18.044069}, our estimates result in a significant mitigation of the damaging effect that intensity correlations have on the secret key rate.

\subsection{Estimation of the conditional photon-number statistics}
\label{Sec stimate n-photon probabilities}

Due to the expression of the correlation parameters ---Eq.~(\ref{equ:tauChongXie})---, 
it is imperative to tightly estimate the conditional photon-number statistics for each possible record of intensity settings. This subsection is devoted to this task.

\begin{figure}[htbp]
    \centering
        \includegraphics[width=0.45 \textwidth]{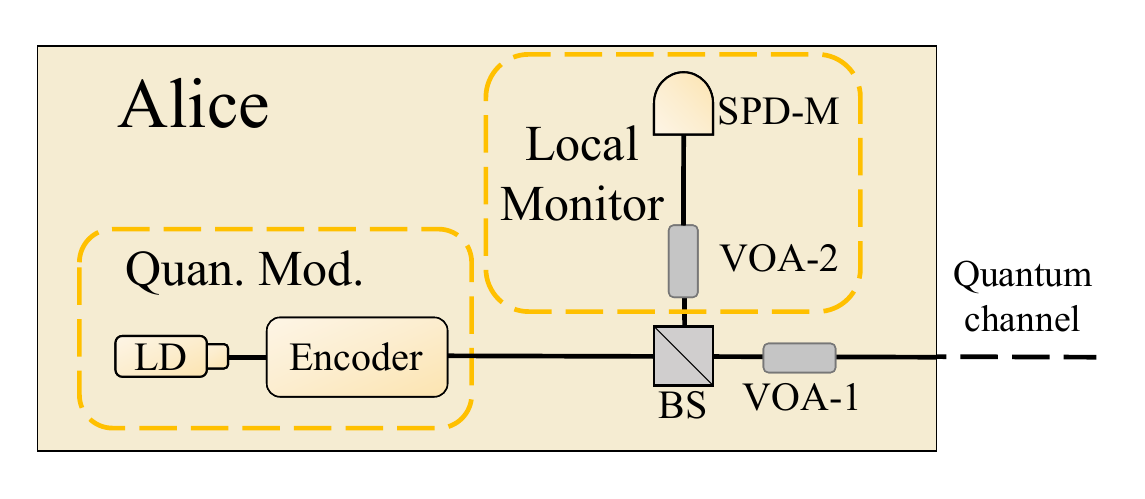}
    \caption{
              Schematic of Alice's module.
               Quan. Mod., quantum module; 
               LD, laser diode;
               Encoder, quantum state and decoy-state intensity modulator;
               BS, beam splitter;
               VOA, variable optical attenuator;
               SPD, single-photon detector.
            }
    \label{fig regulator}
\end{figure}

The working principle of the local monitor is depicted in Fig.~\ref{fig regulator}, although we remark that alternative configurations may fulfil the same purpose as well. By means of a laser diode (LD) and an Encoder, Alice prepares PRWCPs which are then splitted into two components by a balanced BS. While the transmitted pulses enter the quantum channel after undergoing attenuation by a variable optical attenuator (VOA), the reflected pulses are directed to the local monitor, which consists of a VOA (VOA-2) and a single-photon detector (SPD-M). The relative attenuation of the local monitor is defined as 
$\eta_{\text{m}} = {\eta_{\text{d}_{\text{moni}}} \eta_{\text{VOA-1}}}/{\eta_{\text{VOA-2}}}$
where $\eta_{\text{d}_{\text{moni}}}$ is the detection efficiency of SPD-M and
$ \eta_{\text{VOA}-i}$  is the attenuation of $\text{VOA}-i$. As mentioned in the main text, knowing the exact value of the relative attenuation $\eta_{\rm m}$ is crucial for our parameter estimation method.

    The SPD is chosen for several key reasons.  
    First, Alice's local monitoring requires security parameters, 
    specifically intensity correlation parameters, 
    which SPDs provide more accurately due to lower noise than conventional PDs.  
    Second, the threshold response model of SPDs can be used for a tighter estimation of the photon-number statistics,
    which will be shown below.
    Despite limited detection efficiency, SPDs can still accurately measure intensity correlations by grouping click outcomes 
    based on previous round settings and calculating response rates. 

\subsubsection{Bounding the average conditional intensities $\bar{\alpha}_{\{k,\xi\}}$}\label{Sec II}

Combining a standard characterization of single-photon detectors with the general assumptions of Sec.~\ref{sec:Assumptions}, it follows that the detection probability of the local SPD in the $k$-th round conditioned on a record of settings ${\{k,\xi\}}$ is given by
\begin{equation}
    \label{equ:gain1}
D_{\{k,\xi\}}=1-(1-p_d)(1-p_{ap})\int_{\delta_{\{k,\xi\}}^{-}}^{\delta_{\{k,\xi\}}^{+}}g_{\{k,\xi\}}(\delta)\displaystyle{e^{-\eta_{\text{m}} \bar{\alpha}_{\{k,\xi\}}(1+\delta)}}{\rm d}\delta,
\end{equation}
where $p_d$ and $p_{ap}$ respectively denote the dark count rate and the afterpulse rate of the local SPD. Defining $\mathcal{D}_{\{k,\xi\}} = (1-D_{\{k,\xi\}})/[(1-p_d)(1-p_{ap})]$, it follows that
\begin{equation}
    \label{eq lu 2}
\textcolor{black}{\mathcal{D}_{\{k,\xi\}}}
=\int_{\delta_{\{k,\xi\}}^{-}}^{\delta_{\{k,\xi\}}^{+}}  
g_{\{k,\xi\}}(\delta)\displaystyle{e^{-\eta_{\text{m}} \bar{\alpha}_{\{k,\xi\}}(1+\delta)}}{\rm d}\delta
.
\end{equation}
Employing the Taylor expansion ${\rm e}^{x}<1+x+x^2/2$ 
($-1<x<0$) 
in Eq. (\ref{eq lu 2}), we obtain 
\begin{widetext}
\begin{equation}
    \label{equ:Taylorexp_1}
\textcolor{black}{\mathcal{D}_{\{k,\xi\}}}<\int_{\delta_{\{k,\xi\}}^{-}}^{\delta_{\{k,\xi\}}^{+}}{g}_{\{k,\xi\}}(\delta)\left[1-\eta_{\text{m}} \bar{\alpha}_{\{k,\xi\}}(1+\delta)+\frac{\eta_{\text{m}}^2\bar{\alpha}^{2}_{\{k,\xi\}}(1+\delta)^2}2\right]{\rm d}\delta,
\end{equation}
and since the first-order term vanishes (see Eq.~(\ref{equ:note514})), we have that
\begin{equation}
    \label{equ:combing1}
        \begin{aligned}
            \textcolor{black}{\mathcal{D}_{\{k,\xi\}}}
            <
            1-\eta_{\text{m}} \bar{\alpha}_{\{k,\xi\}}
            +\eta_{\text{m}}^2 \bar{\alpha}^{2}_{\{k,\xi\}}
            \frac{
                \zeta_{\{k,\xi\}} + 1
            }2,
        \end{aligned}
\end{equation}
for $\zeta_{\{k,\xi\}} = {\rm max}\{ (\delta_{\{k,\xi\}}^{-})^2 , (\delta_{\{k,\xi\}}^{+})^2 \}$. Solving this quadratic equation for $\bar{\alpha}_{\{k,\xi\}}$ we find the upper bound
\begin{equation}
    \label{equ:SOLVING1}
        \begin{aligned}
            \bar{\alpha}_{\{k,\xi\}}
            < &
            \bar{\alpha}_{\{k,\xi\}}^{\rm U}
            = &
            \frac{
                1-\sqrt{
                    1-2(1-
                    \textcolor{black}{\mathcal{D}_{\{k,\xi\}}}
                    )
                    (1+\zeta_{\{k,\xi\}})
                }
            }{
                {\eta_{\text{m}}} (1+\zeta_{\{k,\xi\}})
            }
        \end{aligned}
        .
\end{equation}
Similarly, keeping the third order we have that ${\rm e}^{x}>1+x+x^2/2+x^3/6$ ($-1<x<0$), 
    and considering the practical situation of QKD,
    we have
    $1 -\eta_m \bar{\alpha}_{\{k,\xi\}} (1 - \frac{\delta}{3})>0$
    when $0<\eta_m , \left\lvert \delta \right\rvert \ll 1$.
    This means that
\begin{equation}
    \label{equ:combing2}
        \begin{aligned}
                \mathcal{D}_{\{k,\xi\}}
            >
            &
                \int_{\delta_{\{k,\xi\}}^{-}}^{\delta_{\{k,\xi\}}^{+}}
                {g}_{\{k,\xi\}}(\delta)
                \left[
                    1-
                    \eta_{\text{m}} \bar{\alpha}_{\{k,\xi\}}(1+\delta)
                    +\frac{\eta_{\text{m}}^2\bar{\alpha}^{2}_{\{k,\xi\}}(1+\delta)^2}2
                    -\frac{\eta_{\text{m}}^3\bar{\alpha}^{3}_{\{k,\xi\}}(1+\delta)^3}6
                \right]{\rm d}\delta
            \\
            > &
            1-{\eta_{\text{m}}} \bar{\alpha}_{\{k,\xi\}}
            +\frac{
                {\eta_{\text{m}}}^2 \bar{\alpha}^{2}_{\{k,\xi\}}
            }2
            -\frac{
                {\eta_{\text{m}}}^3 \bar{\alpha}^{3}_{\{k,\xi\}}
            }{6}.
        \end{aligned}
\end{equation}
\end{widetext}
Coming next, the second order term can be lower-bounded via 
$\bar{\alpha}_{\{k,\xi\}} \geq (1-
\textcolor{black}{\mathcal{D}_{\{k,\xi\}}}
)/{\eta_{\text{m}}}$ 
---which results from plugging the relation ${\rm e}^{x}>1+x$ ($-1<x<0$) in Eq.~(\ref{eq lu 2})
--- and the third order term can be lower bounded via $\bar{\alpha}_{\{k,\xi\}} < \bar{\alpha}_{\{k,\xi\}}^{\rm U}$. 
Altogether, this yields
\begin{equation}
    \label{equ:SOLVING2}
        \begin{aligned}
            \bar{\alpha}_{\{k,\xi\}}
            > &
            \bar{\alpha}_{\{k,\xi\}}^{\rm L}
            \\
            = &
            \frac{1-
            \textcolor{black}{\mathcal{D}_{\{k,\xi\}}}
            }{{\eta_{\text{m}}}}
            +
            \frac{(1-
            \textcolor{black}{\mathcal{D}_{\{k,\xi\}}}
            )^2}{2{\eta_{\text{m}}}}
            -
            \frac{
                {\eta_{\text{m}}}^2(\bar{\alpha}_{\{k,\xi\}}^{\rm U})^3
            }{6}
        \end{aligned}
    .
\end{equation}

\subsubsection{Bounding the conditional photon-number statistics $ p_{n}|_{\{k,\xi\}}$}
The purpose of this subsection is to tightly bound the conditional photon-number statistics, $ p_{n}|_{\{k,\xi\}}$. To this end, we re-use the trick of Taylor-expanding around the conditional expectation $\bar{\alpha}_{\{k,\xi\}}$.
Starting from Eq.~(\ref{equ:pn_all}) and expanding both ${\rm exp}\left({- \bar{\alpha}_{\{k,\xi\}}(1+\delta)}\right)$ and $[\bar{\alpha}_{\{k,\xi\}}(1+\delta)]^n$ in $\delta$ yields
\begin{widetext}
\begin{equation}
    \label{equ:pn_all2equ}
        \begin{aligned}
            p_{n}|_{\{k,\xi\}}
            =
            &
            \frac{
                \bar{\alpha}^{n}_{\{k,\xi\}}
                {\rm e}^{- \bar{\alpha}_{\{k,\xi\}}}
            }{
                n!
            }
            \int_{\delta_{\{k,\xi\}}^{-}}^{\delta_{\{k,\xi\}}^{+}}
            {g}_{\{k,\xi\}}(\delta)
            \left[1-\bar{\alpha}_{\{k,\xi\}}\delta+o(\delta^2)\right]
            \left[1+n\delta+o(\delta^2)\right]{\rm d}\delta
            \\
            =
            &
            \frac{
                \bar{\alpha}^{n}_{\{k,\xi\}}
                {\rm e}^{- \bar{\alpha}_{\{k,\xi\}}}
            }{
                n!
            }
            \int_{\delta_{\{k,\xi\}}^{-}}^{\delta_{\{k,\xi\}}^{+}}
            {g}_{\{k,\xi\}}(\delta)
            \left[1+(n-\bar{\alpha}_{\{k,\xi\}})\delta+o(\delta^2)\right]{\rm d}\delta.
        \end{aligned}
\end{equation}
Since the first-order term $(n-\bar{\alpha}_{\{k,\xi\}})\delta$ vanishes upon integration, one can explicitly suppress it from the definition of $p_{n}|_{\{k,\xi\}}$ to obtain
\begin{equation}\label{equ:pn_all3}
\begin{aligned}
p_{n}|_{\{k,\xi\}}
=
&
p_{n}|_{\{k,\xi\}}-\frac{\bar{\alpha}^{n}_{\{k,\xi\}}{\rm e}^{- \bar{\alpha}_{\{k,\xi\}}}}{n!}\int_{\delta_{\{k,\xi\}}^{-}}^{\delta_{\{k,\xi\}}^{+}}{g}_{\{k,\xi\}}(\delta)(n-\bar{\alpha}_{\{k,\xi\}})\delta{\rm d}\delta\\
=
&
\frac{\bar{\alpha}^{n}_{\{k,\xi\}}{\rm e}^{- \bar{\alpha}_{\{k,\xi\}}}}{n!}\int_{\delta_{\{k,\xi\}}^{-}}^{\delta_{\{k,\xi\}}^{+}}  {g}_{\{k,\xi\}}(\delta)\left[{\rm e}^{-\delta \bar{\alpha}_{\{k,\xi\}}}(1+\delta)^n-(n-\bar{\alpha}_{\{k,\xi\}})\delta\right]{\rm d}\delta 
\\
=
&
\int_{\delta_{\{k,\xi\}}^{-}}^{\delta_{\{k,\xi\}}^{+}}  {g}_{\{k,\xi\}}(\delta)f_n(\delta,\bar{\alpha}_{\{k,\xi\}})  {\rm d}\delta,
\end{aligned}
\end{equation}
for
\begin{equation}
    \label{equ:defineF}
        \begin{aligned}
            f_n(x,y)
            =
            \frac{y^n{\rm e}^{-y}}{n!}
            [
                {\rm e}^{-xy}(1+x)^n
                -(n-y)x
            ]
        \end{aligned}
    .
\end{equation}
This methodology, which is also exploited in~\cite{10050030}, allows for a more accurate estimation of the photon-number statistics than the monotonicity arguments provided in~\cite{zapatero2021security,PhysRevApplied.18.044069}. Particularly, it follows from Eq.~(\ref{equ:pn_all3}) that
\begin{equation}\label{equ:p_nUL}
    \begin{aligned}
\min_{x,y} f_n(x,y)\le{}p_{n}|_{\{k,\xi\}}\le{}\max_{x,y} f_n(x,y),
\\
\quad{\rm for}\quad{}x \in [\delta_{\{k,\xi\}}^{-} , \delta_{\{k,\xi\}}^{+}],\quad{}y \in [\bar{\alpha}_{\{k,\xi\}}^{\rm L} , \bar{\alpha}_{\{k,\xi\}}^{\rm U}].
    \end{aligned}
\end{equation}
On the contrary, invoking the monotonicity of the Poissonian photon-number probabilities in both $\bar{\alpha}_{\{k,\xi\}}$ and $\delta$, it readily follows that
\begin{equation}\label{monotonicity}
    \begin{aligned}
        &
{\rm e}^{-\bar{\alpha}_{\{k,\xi\}}^{\rm L}(1+\delta_{\{k,\xi\}}^{-})}\frac{\left[\bar{\alpha}_{\{k,\xi\}}^{\rm L}\left(1+\delta_{\{k,\xi\}}^{-}\right)\right]^n}{n!}
\le{}
p_{n}|_{\{k,\xi\}}
\le{}
{\rm e}^{-\bar{\alpha}_{\{k,\xi\}}^{\rm U}(1+\delta_{\{k,\xi\}}^{+})}\frac{\left[\bar{\alpha}_{\{k,\xi\}}^{\rm U}\left(1+\delta_{\{k,\xi\}}^{+}\right)\right]^n}{n!},
    \end{aligned}
\end{equation}
for $n\geq{}1$. This latter approach overlooks the fact that the first-order term in $\delta$ vanishes upon integration, resulting in looser bounds than those of Eq.~(\ref{equ:p_nUL}).

Since the tighter bounds of Eq.~(\ref{equ:p_nUL}) further rely on explicit numerical optimization, we use these bounds only up to a certain threshold photon number $n_{\rm th}$, and for $n>n_{\rm th}$ we apply the monotonicity bounds of Eq.~(\ref{monotonicity}) directly. In either case, we refer to these lower and upper bounds as $p_{n}|_{\{k,\xi\}}^{\rm L}$ and $p_{n}|_{\{k,\xi\}}^{\rm U}$, respectively.

\subsection{Estimation of the correlation parameters}
\label{sec asdsfhakscfaeracb}
Let us now calculate the lower bounds on the correlation parameters ---defined in Eq.~(\ref{equ:tauChongXie})--- that arise from the $p_{n}|_{\{k,\xi\}}^{\rm L}$ of the previous subsection. Splitting $\sum_{m=0}^{\infty}\sqrt{p_{m}^{\rm L}|_{a_{i-\xi}\ldots{}a_{k}\ldots{}a_{i}}p_{m}^{\rm L}|_{a_{i-\xi}\ldots{}a_{k}'\ldots{}a_{i}}}$ into two sums to separately make use of Eq.~(\ref{equ:p_nUL}) and Eq.~(\ref{monotonicity}) one can readily show that
\begin{equation}\label{bounds_on_tau}
\begin{aligned}
&
\sum_{m=0}^{\infty}\sqrt{p_{m}^{\rm L}|_{a_{i-\xi}\ldots{}a_{k}\ldots{}a_{i}}p_{m}^{\rm L}|_{a_{i-\xi}\ldots{}a_{k}'\ldots{}a_{i}}}\\
&
        \geq{}\sum_{m=0}^{n_{\rm th}}
                    \sqrt{
                        \mathop{\rm min}\limits_{x_1,y_1}f_m(x_1,y_1)
                        \mathop{\rm min}\limits_{x_2,y_2}f_m(x_2,y_2)
                    }
        +
        \sum_{m=n_{{\rm th}+1}}^{\infty}
        {\rm e}^{-a_k\left(1-\delta_{{\rm max},a_k}^{\rm rand}\right)\left(1-\delta_{{\rm max},a_k}^{\rm corr}\right)}
        \frac{
        \left[
        a_k\left(1-\delta_{{\rm max},a_k}^{\rm rand}\right)\left(1-\delta_{{\rm max},a_k}^{\rm corr}\right)
        \right]^m
        }{m!}
        \\
        =
        &
        1
        +
        \sum_{m=0}^{n_{\rm th}}
        \left[
                    \sqrt{
                        \mathop{\rm min}\limits_{x_1,y_1}f_n(x_1,y_1)
                        \mathop{\rm min}\limits_{x_2,y_2}f_n(x_2,y_2)
                    }
        -
        {\rm e}^{-a_k^{\rm L}}
        \frac{
        (
        a_k^{\rm L}
        )^n
        }{n!}
        \right]
            ,
    \end{aligned}
\end{equation}
where
$x(y)_i \in [ x(y)_i^{\rm L},x(y)_i^{\rm U} ]$,
$x_1^{\rm L} = \delta_{a_{i-\xi}\ldots{}a_{k}\ldots{}a_{i}}^{-}$,
$x_1^{\rm U} = \delta_{a_{i-\xi}\ldots{}a_{k}\ldots{}a_{i}}^{+}$,
$x_2^{\rm L} = \delta_{a_{i-\xi}\ldots{}a_{k}'\ldots{}a_{i}}^{-}$,
$x_2^{\rm U} = \delta_{a_{i-\xi}\ldots{}a_{k}'\ldots{}a_{i}}^{+}$,
$y_1^{\rm L} = \bar{\alpha}_{a_{i-\xi}\ldots{}a_{k}\ldots{}a_{i}}^{\rm L}$,
$y_1^{\rm U} = \bar{\alpha}_{a_{i-\xi}\ldots{}a_{k}\ldots{}a_{i}}^{\rm U}$,
$y_2^{\rm L} = \bar{\alpha}_{a_{i-\xi}\ldots{}a_{k}'\ldots{}a_{i}}^{\rm L}$,
$y_2^{\rm U} = \bar{\alpha}_{a_{i-\xi}\ldots{}a_{k}'\ldots{}a_{i}}^{\rm U}$
and $a_k^{\rm L} = a_k(1-\delta_{{\rm max},a_k}^{\rm rand})(1-\delta_{{\rm max},a_k}^{\rm corr})$. Plugging this into Eq.~(\ref{equ:tauChongXie}) yields
\begin{equation}
        \label{equ:tauneq}
        \begin{aligned}
            \tau^{\xi}_{ a_{k}, a_{k}' , n}
            \geq
            &
            \left\{
            \sum_{a_{k+1}\in A}
                \ldots
                \sum_{a_{k+\xi}\in A}
                \left[
                    \left(
                        \prod_{i=k+1}^{k+\xi} p_{a_i}
                    \right)
                    \left\{
                        1
        +
        \sum_{n=0}^{n_{\rm th}}
        \left[
                    \sqrt{
                        \mathop{\rm min}\limits_{x_1,y_1}f_n(x_1,y_1)
                        \mathop{\rm min}\limits_{x_2,y_2}f_n(x_2,y_2)
                    }
        -
        {\rm e}^{-a_k^{\rm L}}
        \frac{
        (
        a_k^{\rm L}
        )^n
        }{n!}
        \right]
        \right\}
        \right]
        \right\}
                .
        \end{aligned}
\end{equation}
\end{widetext}

    However,
    Although the above analysis assumes that 
    $\eta_m$
  can be accurately characterized, if calibration errors of 
$\eta_m$
  do occur, we can still analyze and estimate the conditional photon-number statistics and correlation parameters. If the range of 
$\eta_m$
  can be determined, it can be incorporated into Eq. (\ref{equ:SOLVING1}) and (\ref{equ:SOLVING2}) to estimate new upper and lower bounds for 
$\bar{\alpha}_{\{k,\xi\}}$. 
By combining this with Eq. (\ref{equ:tauneq}), we can obtain updated intensity correlation parameters. 
In subsequent analysis, calculations would be based on the worst-case result within the range of 
$\eta_m$.
Additionally, we have numerically verified that an exact characterization of $\eta_m$ is not essential. 
A relative deviation in $\eta_m$ only results in a similarly scaled relative deviation in the secret key rate.

\section{Asymptotic secret key rate and simulation results}\label{keyrate}

\begin{figure*}
    \centering
      \subfigure[{Secret key rate when Bob uses single-photon detectors (SPDs) with detection efficiency $\eta_{\rm det} = 0.2$ and dark count rate $p_d = 4.2 \times 10^{-6}$.}]{
        \label{fig: SPDrate}
        \includegraphics[width=0.48\textwidth]{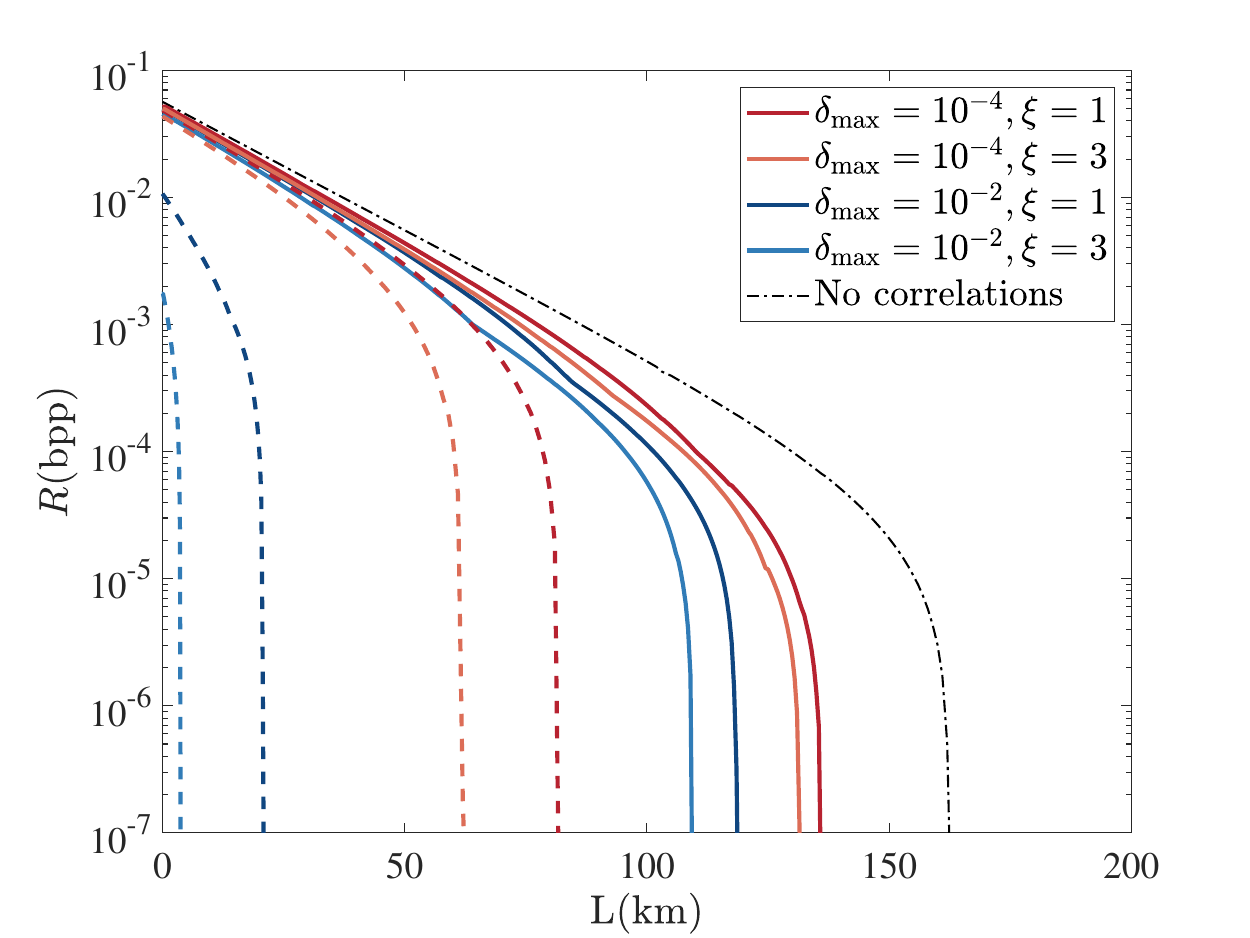}
        }
      \subfigure[{Secret key rate when Bob uses superconducting nanowire SPDs (SNSPDs) with detection efficiency $\eta_{\rm det} = 0.608$ and dark count rate $p_d = 9.5 \times 10^{-8}$.}]{
        \label{fig: SSPDrate}
        \includegraphics[width=0.48\textwidth]{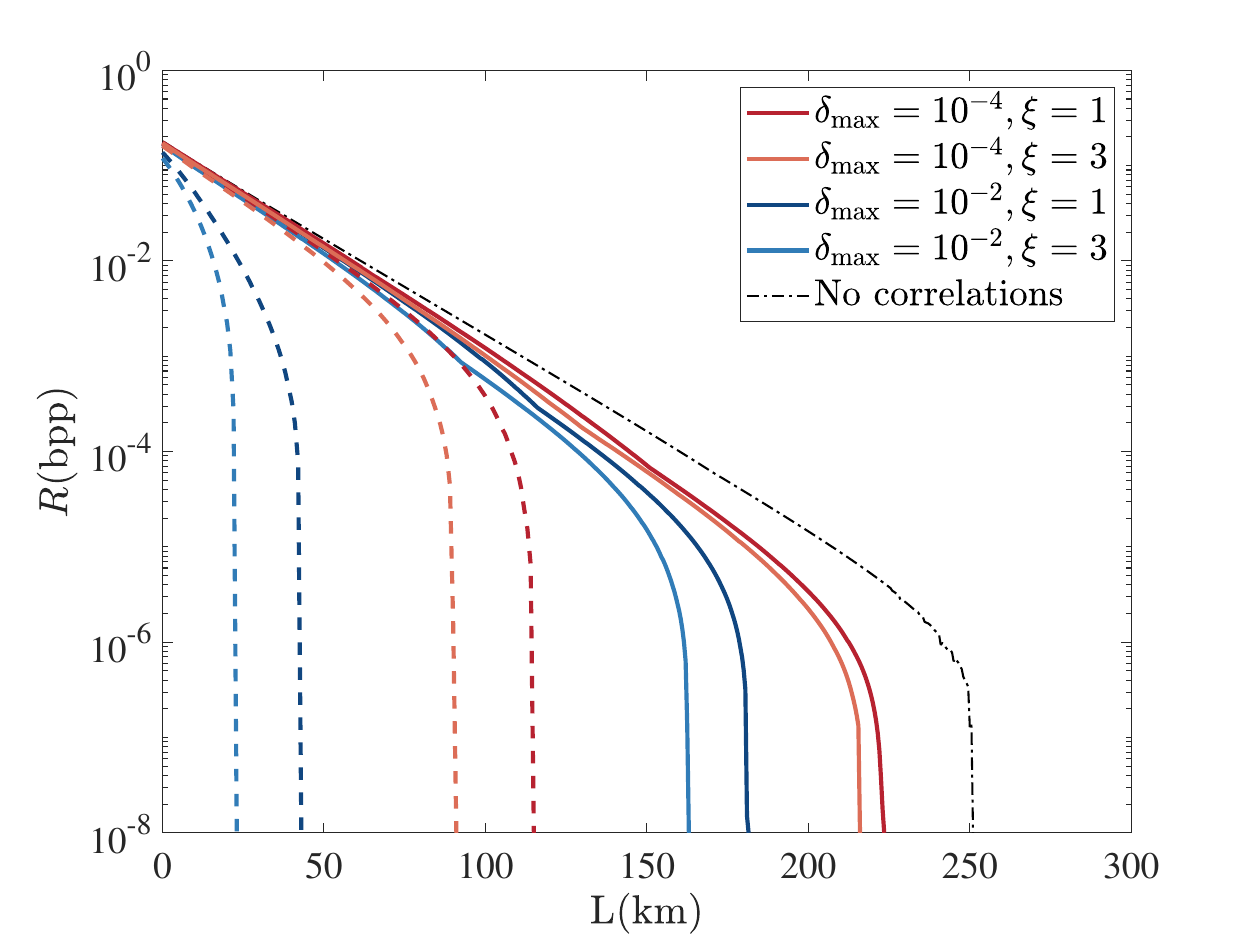}
        }
    \caption{
               Simulation of the asymptotic secret key rate \textcolor{black}{in an asymmetric protocol with an active receiver. Solid lines, secret key rate obtained with our method; 
               Dashed lines, secret key rate obtained with the most efficient previous method \cite{PhysRevApplied.18.044069}. 
               We set $\delta_ {{\rm max},a_k} ^ {\rm corr} = \delta_ {{\rm max},a_k} ^ {\rm rand} = {\delta_{\rm max}}/2$ for $\delta_ {\rm max} \in \{10 ^ {-2}, 10 ^ {-4} \} $, 
               and contemplate two correlation ranges, $\xi \in \{1, 3 \} $. We focus on the (asymptotically optimal) limit where $p_ {\mu} = 1, p_ {\nu} = 0, p_ {\omega} = 0 $ and $q_Z = 1$,} 
               {where in this case $q_Z$ determines Bob's probability to actively select the $Z$-basis as well.} \textcolor{black}{Also, we set the fiber attenuation coefficient to $\alpha = 0.2$ dB/km, the 
               misalignment to $\delta_A=0.08$ and the error correction efficiency to $f_{\rm EC}=1.16$. 
               }}
      \label{fig:keyRate}
\end{figure*}

In the simulations, we follow the asymptotic analysis. 
For large enough $N$, one can identify the round-averaged observables with their expectations to approximate the secret key rate via
\begin{equation}
    \label{equ:asymptotic_secret_key_rate}
        \begin{aligned}
            R
            \approx
            Z_{1,\mu}^{\rm L}
            \left[
                1
                -H\left(
                    \frac{
                        E_{1,\mu}^{\rm U}
                    }{
                        X_{1,\mu}^{\rm L}
                    }
                \right)
            \right]
            -f_{\rm EC}
            Z_{\mu}
            H(E_{\rm tol})
        \end{aligned}
    ,
\end{equation}
where $Z_{1,\mu}^{\rm L}=\sum_{k=0}^{N}Z_{1,\mu}^{(k)\mathrm{L}}/N$, $X_{1,\mu}^{\rm L}=\sum_{k=0}^{N}X_{1,\mu}^{(k)\mathrm{L}}/N$, $E_{1,\mu}^{\rm U}=\sum_{k=0}^{N}E_{1,\mu}^{(k)\mathrm{U}}/N$, $f_{\rm EC}$ is the error-correction efficiency, $H(x)$ is the binary entropy function, $E_{\rm tol}$ is the expected bit error rate in the $Z$-basis, and $Z_{\mu}=\sum_{k=0}^{N}Z_{\mu}^{(k)}/N$ for
\begin{equation}\label{round-dependent-signal-gain}
Z_{\mu}^{(k)}=\sum_{a_{k-\xi}\in{}A}\ldots{}\sum_{a_{k-1}\in{}A}Z_{\{k-1,\xi-1\}\mu}.
\end{equation}
Note that $Z_{\mu}$ is the expected ratio of rounds where both parties select the $Z$-basis, Alice selects the signal setting $\mu$, and a detection event is registered. Also, we remark that $Z_{1,\mu}^{\rm L}$ incorporates the sifting factor $q_{\rm Z}^2$ by definition (see Eq.~(\ref{target_yields})). Of course, alternatively, one could consider a symmetric scenario where the parties extract key from both bases.

In the numerical simulations, we use the same channel and detector model deployed in~\cite{zapatero2021security,PhysRevApplied.18.044069}, which is consistent with the reference parameters derived in Sec.~\ref{sec:sdsfvsdaa}. For ease of comparison with prior work, this model does not incorporate intensity correlations. The parameters of the model are $\eta_{\rm det}$ (detector efficiency of Bob's detectors), $p_d$ (dark count probability of Bob's detectors), $\eta_{\rm ch} = 10^{-\alpha_{\rm att} L/10}$ (channel transmittance for a lab-to-lab distance of $L$ kilometers and an attenuation coefficient of $\alpha_{\rm att}$ dB/km) and $\delta_A$ (misalignment error in the channel). The model reads
\begin{equation}
    \label{equ:1145}
        \begin{aligned}
        &
            \frac{
                    Z_{\{k,\xi\}}
            }{
                q_Z^2 \prod_{i = k-\xi}^{k} p_{a_i}
            }
            =
            \frac{
                    X_{\{k,\xi\}}
            }{
                q_X^2 \prod_{i = k-\xi}^{k} p_{a_i}
            }
            =
            1
            -(1-p_d)^2
            {\rm e}^{-\eta a_k},\\
            &
             \frac{
                    E_{\{k,\xi\}}
            }{
                q_X^2 \prod_{i = k-\xi}^{k} p_{a_i}
            }
            =
            \frac{p_d^2}2
            +
            p_d
            (1-p_d)
            \left[1+h_{\eta,\delta_{A}}(a_{k})\right]
            \\ & \quad
            \quad \quad \quad \quad \quad \quad
            +
            (1-p_d)^2
            \left[\frac12+h_{\eta,\delta_{A}}(a_{k})-\frac12 {\rm e}^{-\eta a_k}\right],\\
        \end{aligned}
\end{equation}
for $\eta = \eta_{\rm ch}\eta_{\rm det}$ and $h_{\eta,\delta_{A}}(a_{k})=\bigl({\rm e}^{-\eta a_k \cos^2\delta_A}-{\rm e}^{-\eta a_k \sin^2\delta_A}\bigr)/2$. Naturally as well, $E_{\rm tol}={E_{\{k,\xi\}}}/{X_{\{k,\xi\}}}$ because of the symmetry between the $Z$ and the $X$ bases.

{In Fig.~\ref{fig:keyRate} we present numerical simulations of the asymptotic secret key rate attainable with our enhanced decoy-state method. For ease of comparison with prior work~\cite{zapatero2021security,PhysRevApplied.18.044069}, we consider an asymmetric decoy-state BB84 protocol with an active receiver, using the key rate formula and the channel and detector model provided in Sec.~\uppercase\expandafter{\romannumeral4} of the Supplemental Material.} The figure reveals that this method offers a great tolerance to variations in the magnitude and range of the correlations when compared to previous studies. In fact, it consistently achieves higher secret key rates for all considered scenarios.

{Note that, in this work, we use the model here presented for the simulations of both Fig.~\ref{fig:keyRate} and Fig.~\ref{fig:EXPRFIG}. In this respect, we remark that, although the model is specifically suited to an active BB84 receiver, it can also be used to approximately describe the detection statistics of a passive receiver.}

\section{Experimental Setup and Implementation}

\textcolor{black}{The QKD setup consists} of three modules. On the source side, as usual, there is a quantum module \textcolor{black}{responsible for the BB84 and decoy-state encoding}. Subsequently, the monitor module \textcolor{black}{monitors} the quantum module, capturing the intensity correlation information. The third module is Bob's detection setup. Precisely, in the quantum module, \textcolor{black}{for each protocol round $k = 1, 2, \ldots, N$, Alice selects an intensity setting $a_k \in A = \{\mu, \nu, \omega\}$ with probability $p_{a_k}$, and a bit-and-basis setting $r_k \in R = \{Z0, Z1 , X0, X1\}$ with probability $p_{r_k}$. The module encodes these settings on PRWCPs that are attenuated by a beam splitter (BS) and a variable optical attenuator (VOA). Due to the effect of intensity correlations, the average photon-number of the $k$-th PRWCP is not exactly $a_k$, but $\alpha_{k}$ according to the general model described above.}

\textcolor{black}{The output of the quantum module consists of two parts, the first one being sent to Bob and the second one being sent to the monitor module. The monitor module contains a VOA and an SPD, 
and it classifies the detection results according to the record of the previous $\xi$ settings, for the later calculation of (i) the bounds of the conditional $n$-photon probabilities of the different records, 
and (ii) the correlation parameters $\tau^{\xi}_{ a_{k}, a_{k}' , n}$. For this purpose, the method described in 
Sec.~\ref{sec:asdcqwecsg333}
is used.
}

\textcolor{black}{In the detection module, Bob measures the received signals and announces whether a successful result occurs in his measurement unit (MU). He stores the rounds of successful measurements and the corresponding bases and raw key bits. For successful measurements, Alice stores her original key bit and basis, while she stores her decoy settings for all rounds to recognize the records $\{k,\xi\}$ of every successful measurement.} Box~\ref{bx:ict-1} illustrates the steps of the protocol,  where the additional operations introduced to address the intensity correlation problem are highlighted in bold.

\begin{figure*}[htbp]
    \centering
        \includegraphics[width=0.8\textwidth]{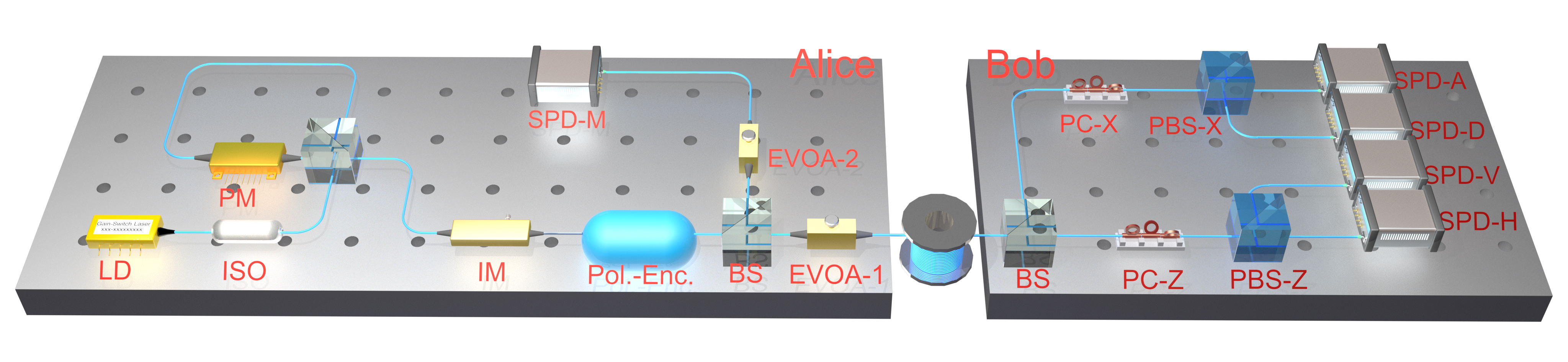}
    \caption{
               Depiction of the experimental setup.
               LD, laser diode;
               ISO, isolator;
               PM, phase modulator;
               IM, intensity modulator;
               Pol.-Enc., polarization encoder;
               BS, beam splitter;
               EVOA, electronic variable optical attenuator;
               SPD, single-photon detector;
               PC, polarization controller;
               PBS, polarization beam splitter.
            }
            \label{fig:EXPFIG}
\end{figure*}

To validate our approach, we conducted experiments using the apparatus shown in Fig. \ref{fig:EXPFIG}. 
We use a polarization coding system where a gain-switched laser diode emits PRWCPs with a pulse width of 50 ps and a repetition rate of 1 GHz. 
The decoy-state intensity modulator (DS-IM) module includes an isolator, a Sagnac interferometer (SI) \cite{Roberts:18}, 
and a commercial IM made from an integrated ${\rm LiNbO_3}$ Mach-Zehnder interferometer (MZI). 
The SI employs a phase modulator (PM) to modulate clockwise and counterclockwise pulses, 
causing interference that leads to maximum and minimum attenuation when the relative phase is $0$ and $\pi$, respectively. 
Due to the SI's unique characteristics \cite{Roberts:18}, it has two stable operating points during constructive and destructive interference, 
achieving smaller fluctuations and correlations compared to the commercial IM. 
SI is used to adjust the signal and decoy state, while the vacuum state is modulated using MZI, 
which also has stable operating points during constructive and destructive interference \cite{Yoshino2018,9907821}. 
By combining the SI and MZI IM, each with two stable operating points at maximum and minimum attenuation, 
we achieve the three required stable operating points for the DS-IM module. The specific interference combinations are shown in Tab. \ref{tab:1}.

\begin{table}[h]
    \centering
    \caption{
        \textcolor{black}{Relation between the 
        intensity settings and the state of the interferometers}. 
        Cons., 
        constructive interference;
        Des., 
        destructive interference.
    }
    \label{tab:1}
    \begin{ruledtabular}
    \begin{tabular}{c|ccc}
     & Signal & Decoy & Vacuum \\ \hline
    SI       & Cons. & Des. & Des. \\
    MZI      & Cons. & Cons.  & Des. 
    \end{tabular}
    \end{ruledtabular}
\end{table}

\textcolor{black}{The polarization encoder encodes the DS-IM modulated pulses. 
A BS then divides the encoded states into two pulses,
one of them entering the local monitor ---composed of an 
electronic variable optical attenuator (EVOA-2) and a single-photon detector (SPD-M)--- 
and the other one entering the fiber-based quantum channel after being attenuated with EVOA-1. 
}

{At the detection side, we opt for a passive BB84 receiver to measure the arriving polarization states.} The detection module comprises one \textcolor{black}{symmetric} BS and two MUs. Each MU consists of a polarization controller (PC), a polarization BS (PBS), and two SPDs. 
The SPDs work on gated mode \cite{10.1063/1.4978599} at a frequency of 1 GHz, 
with a dark count rate $p_d=4.2\times 10^{-6}$
and a detection efficiency $\eta_{\rm det}=20\%$.
The PC-$Z$ of MU-$Z$ is properly adjusted so that SPD-H (V) measures horizontal (vertical) polarization states.  
\textcolor{black}{In a similar fashion, MU-$X$ measures the $X$ basis polarization states.}

The attenuation of Alice's local monitor is controlled precisely to achieve the desired value of $\eta_{\text{m}}$. In our experiment, we set $\eta_{\text{m}}$ to $10^{-3}$. SPD-M records whether or not each round clicks and categorizes all data into $3^{\xi+1}$ types based on the record of settings. \textcolor{black}{To illustrate our method, we have assumed that $\xi = 3$ \cite{9907821} for the data analysis, thus calculating the detection rates of 81 records of settings in total.}

Before performing the experiment, we conduct simulations to determine the intensities $\mu$, $\nu$  and $\omega$ and their selection probabilities.
We adjust the splitting ratio of the BS within SI to fix the signal to decoy intensities
constrained to $\mu/\nu = 5$.
We optimize both the intensities and their probabilities simultaneously in the simulation, 
while imposing the constraint $p_{a_k} \geq 0.15$ for all $a_k$ \textcolor{black}{to ensure that all data sets are sufficiently large. 
For consistency with the experiment, Alice's $Z$-basis probability is set to $q_Z=0.5$ and a symmetric passive receiver is considered at Bob's side. 
In addition, we estimate} 
the magnitude of the intensity correlations in advance as
$\delta_ {{\rm max},\mu}^{\rm corr} = 3.0 \times 10^{-3}$,
$\delta_ {{\rm max},\nu}^{\rm corr} = 2.0 \times 10^{-3}$,
$\delta_ {{\rm max},\omega}^{\rm corr} = 1.6 \times 10^{-2}$ and
$\delta_ {{\rm max},\mu(\nu,\omega)}^{\rm rand} = 3.0 \times 10^{-2}$, \textcolor{black}{and we set the remaining parameters as in Fig. \ref{fig:keyRate}}. Notably, prior knowledge of the device's correlations and fluctuations is not essential for implementing the protocol, but it allows to optimize the secret key rate for each distance. Furthermore, this knowledge can be acquired by Alice's monitor module before initiating the quantum communication.

\begin{figure}[htbp]
    \centering
        \includegraphics[width=0.48\textwidth]{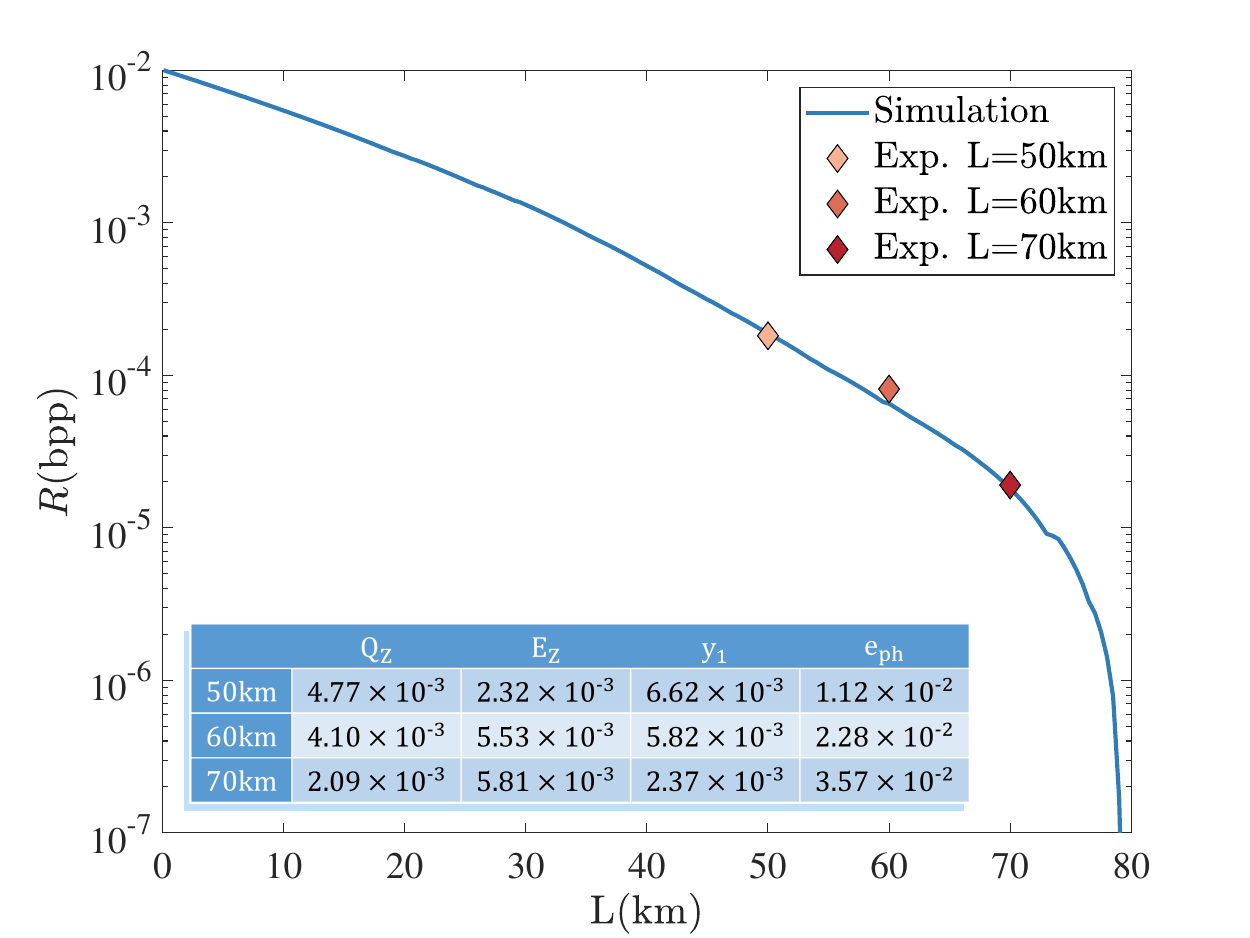}
    \caption{
                {Simulation and experimental performance of the intensity-correlation-tolerant QKD protocol in Box~\ref{bx:ict-1}.} 
                The table inside the figure provides the key parameters of the experiments,
                where
                $Q_Z$, $E_Z$, $y_1$ and $e_{\rm ph}$
                correspond, respectively, to the $Z$ basis gain, the 
                $Z$ basis QBER, the
                $Z$ basis single-photon yield and the phase error rate.
            }
            \label{fig:EXPRFIG}
\end{figure}

\textcolor{black}{The results of the simulation are shown in Fig.~\ref{fig:EXPRFIG}.} We successfully demonstrate the feasibility of our approach at 50 km, 60 km, and 70 km of channel length. As depicted in the figure, 
the \textcolor{black}{asymptotic key rates are respectively given by}
$1.82\times 10^{-4} $,
$8.13\times 10^{-5} $ and
$1.91\times 10^{-5} $, 
{suitably aligned with the corresponding simulation outcomes 
}
{
    (see the Supplemental Material for detail data).
}

\section{CONCLUSIONS}

In summary, we have proposed an efficient solution to the intensity correlation problem in practical QKD. Our approach achieves high tolerance to this imperfection through the addition of a monitor module at Alice's side at minimal cost, leading to higher secret key rates and extended maximum achievable distances compared to previous proposals. Leveraging a stable DS-IM module, we have conducted the first experiment addressing intensity correlations in a practical QKD setup \textcolor{black}{without requiring hardware-based suppression methods~\cite{9907821}, or limiting software-based methods restricted to the nearest-neighbours scenario~\cite{Yoshino2018}. Despite we have considered a finite correlation range, our results could be promoted to the unbounded range setting using the tools in~\cite{pereira2024quantum}.}
{
    For future work, analyzing potential security vulnerabilities arising from the source monitoring module would be valuable.
}

Given that this type of correlations pose a considerable security concern in QKD systems, our work constitutes a significant advance for the practical security of QKD and facilitates the adoption of QKD in large-scale security applications.


\section*{funding.}

The National Natural Science Foundation of China (Grant No. 62271463, 62301524, 62105318, 61961136004, 62171424), the Fundamental Research Funds for the Central Universities, the China Postdoctoral Science Foundation (Grant No. 2022M723064), Natural Science Foundation of Anhui (No. 2308085QF216), and the Innovation Program for Quantum Science and Technology (Grant No. 2021ZD0300700). VZ and MC acknowledge support from the Galician Regional Government (consolidation of Research Units: AtlantTIC), the Spanish Ministry of Economy and Competitiveness (MINECO), the Fondo Europeo de Desarrollo Regional (FEDER) through the grant No. PID2020-118178RB-C21, MICIN with funding from the European Union NextGenerationEU (PRTR-C17.I1) and the Galician Regional Government with own funding through the “Planes Complementarios de I+D+I con las Comunidades Autónomas” in Quantum Communication, the European Union’s Horizon Europe Framework Programme under the Marie Sklodowska-Curie Grant No. 101072637 (Project QSI) and the project “Quantum Security Networks Partnership” (QSNP, grant agreement No. 101114043).

\section*{acknowledgments.}

\section*{Disclosures.}

The authors declare no conflicts of interest.

\section*{Data availability.}

Data underlying the results presented in this paper are not publicly available at this time but may be obtained from the authors upon reason-able request.

\section*{Supplemental document.}

See Supplemental Material for supporting content.


\bibliographystyle{apsrev4-1}
\bibliography{1}

\end{document}